## НАНОЭЛЕКТРОННЫЕ ПРИБОРЫ

# ТУННЕЛЬНЫЙ ПОЛЕВОЙ ТРАНЗИСТОР НА КВАНТОВОЙ ЯМЕ, ОБРАЗОВАННОЙ ТРЕМЯ СЛОЯМИ ПРОИЗВОДНЫХ ГРАФЕНА $(COH)_n$–$(CF)_n$–$(CH)_n$, И СО СТОКОМ ИЗ СРЕДНЕГО СЛОЯ – $(CF)_n$

© 2014 г.  В. А. Жуков[1], В. Г. Маслов[2]

[1]*Санкт-Петербургский институт информатики и автоматизации Российской АН*
[2]*Санкт-Петербургский Государственный университет информационных технологий, механики и оптики*
E-mail: valery.zhukov2@gmail.com
E-mail: maslov04@bk.ru

Рассмотрен вариант туннельного полевого нанотранзистора на квантовой яме, в котором управляющее напряжение 0.6 В прикладывается к окружающим яму барьерам, а сток электронов происходит из квантовой ямы. Электроны туннелируют в квантовую яму через первую половину двугорбого туннельного барьера (двойного гетероперехода), образованного сэндвичем из трех слоев широко зонных 2D полупроводников (производных графена: пергидроксиграфен $(COH)_n$, флюорографен $(CF)_n$, графан $(CH)_n$ с резко отличающимися уровнями дна зоны проводимости. Средний слой – флюорографен $(CF)_n$ имеет низшее дно зоны проводимости, которое и образует квантовую яму глубиной ~3 эВ и шириной ~0.6 нм в общем туннельном потенциальном барьере шириной 1.8 нм и служит каналом для стока электронов. Истоковый и затворный металлические электроды прилегают к внешним слоям 2D полупроводников сэндвича – пергидрооксиграфену $(COH)_n$ и графану $(CH)_n$, соответственно, образуя общий "затворный" сэндвич, размером $20 \times 20$ нм². Металлический стоковый электрод с шириной 10 нм и с потенциалом на 1 В выше, чем у первого электрода (истока) прилегает к выходящему за пределы сэндвича среднему слою флюорографена $(CF)_n$, имеющему для этого большую 35 нм ширину, чем внешние слои 2D полупроводников внутреннего трехслойного сэндвича. Потенциал открывания затвора равен 0.62 В. Максимальный рабочий ток $I_{sd} = 2 \times 10^{-5}$ А. Ток со стока в закрытом состоянии равен нулю, а паразитный ток, текущий через затворный электрод в открытом состоянии $I_g = I_{leak} \sim 10^{-10}$ А. Квантовая емкость транзистора позволяет устройству работать при частоте до $10^{12}$ Гц.

DOI: 10.7868/S0544126914060106

## ВВЕДЕНИЕ

В последние годы появился ряд работ, в которых рассматривается возможность построения нанотранзисторов терагерцового диапазона, как на кремнии [1, 2], так и на графене и его производных [3–5]. В недавней работе Нобелевского лауреата А. Гейма с сотрудниками [6] предложена идея более широкого использования Ван дер Ваальсовских гетероструктур, сформированных слоями широкозонных 2D полупроводников для создания новых устройств наноэлектроники. В работе [6] предлагается заранее выращивать монослойные пленки из широкозонных полупроводников типа флюорографен $(CF)_n$, гексагональный нитрид бора (hexagonal BN), дисульфид вольфрама $(WS_2)$, диселенид молибдена $(MoSe_2)$ и т.д. на подходящих для каждого такого 2D полупроводника подложках. Затем последовательно настилать такие пленки на полированную монокристаллическую подложку, например, из кремния сапфира или кварца. После этого предполагается создавать на получившейся многослойной структуре рельефный рисунок (топографию нужной интегральной схемы), методами нанолитографии. В работе [6] высказывается мнение, что предложенный метод при переходе на технологию "22 нм" и далее будет гораздо производительнее, надежнее и точнее, чем используемый в настоящее время для получения гетероструктур метод молекулярно-пучковой эпитаксии.

В работах [1–5] рассматривались как вертикальные, так и планарные конфигурации нанотранзисторов, использующие только один туннельный барьер.

Следует отметить, что еще в 1977 году в работе Л.Л. Чанга и Л. Эсаки [7] (см. также [8]) была предложена идея построения транзистора с туннельно-тонкой базой, состоящего из двух гетеропереходов, образованных тремя слоями разных полупроводников. Благодаря такой конструкции образуется резкий и одновременно очень узкий провал в зонной диаграмме сэндвича построенного из трех полупроводниковых слоев. Однако ширина этого провала была еще достаточна, для

Результат расчета энергетических зон призводных графена

| Материал | $(COH)_n$ | $(CF)_n$ | $(CH)_n$ |
|---|---|---|---|
| Параметр | $E$ (эВ) | $E$ (эВ) | $E$ (эВ) |
| Дно зоны проводимости $E_{cb}$ | −1.823 | −4.435 | −0.952 |
| Потолок валентной зоны $E_{vb}$ | −4.985 | −7.728 | −4.762 |
| Ширина запрещенной зоны | $\Delta E_{bg} = 3.162$ | $\Delta E_{bg} = 3.279$ | $\Delta E_{bg} = 3.728$ |

того, чтобы область перехода можно было считать трехмерной. В такой конструкции средний слой полупроводника служил туннельным переходом и одновременно являлся базой (затвором). Авторы ожидали получить в таком транзисторе значительное увеличение туннельного тока, но он не был реализован из-за несовершенства технологии, существовавшей на тот момент.

Спустя десятилетие в 1988 г., в работе [9] была предложена уже двух барьерная конфигурация туннельного нанотриода, в которой область между барьерами предполагалась достаточно тонкой, чтобы считать ее двумерной квантовой ямой. В такой конструкции уже предполагалось, что сток осуществляется из области между барьерами, т.е. из квантовой ямы, а внешние по отношению к обоим барьерам электроды, служат истоком и затвором, соответственно.

В этой работе [9] было также впервые введено понятие квантовой емкости двумерного электронного газа (2DEG) в квантовой яме и качественно оценено влияние этой емкости на быстродействие системы в целом. Реализовать подобную конфигурацию в эксперименте в то время также не удалось из-за технологических трудностей. Сравнительно недавно [10] в эксперименте была реализована конструкция транзистора микроволнового радиочастотного диапазона на квантовой яме, образованной δ-легированным слоем кремния, размещенным внутри слоистой гетероструктуры созданной с помощью молекулярно-пучковой эпитаксии и со слоями толщиной 15–30 нм.

В настоящей работе мы рассмотрим конструкцию туннельного нанотранзистора со стоком из квантовой ямы, предложенную в [9], использовав при этом идею Гейма [6] о применении широкозонных 2D полупроводников типа производных графена для построения такого нано устройства. Рассматриваемая конструкция транзистора является "гибридной", т.е. вертикальной по отношению истока к затвору и планарной в отношении истока к стоку. Поэтому мы надеемся воспользоваться преимуществами обеих схем: компактностью вертикальной и быстродействием и возможностью достигать насыщения тока открытого состояния планарной конструкции.

По сравнению с работой [9] и работой [10], в нашей модели предполагается использовать меньшую толщину всех 3 = x слоев сэндвича (~0.6 нм вместо 15–30 нм в [10] или 3–6 нм в [9]). Последнее обстоятельство позволяет ожидать значительно больший туннельный ток на выходе устройства.

## 1. ПОСТАНОВКА ЗАДАЧИ. КВАНТОВО-ХИМИЧЕСКИЙ РАСЧЕТ ШИРОКОЗОННЫХ 2D ПОЛУПРОВОДНИКОВ. ВОЗМОЖНОСТЬ СОЗДАНИЯ НА НИХ НАНОТРАНЗИСТОРА

В нашей работе мы решили сосредоточить внимание на трех материалах — производных графена. Это пергидрокси графен, представляющий собой графен с присоединенными с обеих сторон группами гидроксила, имеющий среднюю структурную химическую формулу $(COH)_n$, флюорографен $(CF)_n$ и графан $(CH)_n$. Были проведены уточненные квантово-химические расчеты зонной структуры этих широко-зонных 2D полупроводников методом DFT (GGA-bp87) в базисе TZP [11, 12] с оптимизацией геометрии, как элементарной ячейки, так и параметров 2D-решетки. Получены значения уровней энергии для дна зоны проводимости $E_{cb}$ и потолка валентной зоны $E_{vb}$ или в терминологии квантовой химии Elumo (Lowest Unoccupied Molecular Orbital) и Ehomo (Highest Occupied Molecular Orbital), соответственно, для трех выбранных материалов.

Из таблицы видно, что ширина запрещенной зоны, т.е. величина разности $\Delta E_{bg}$ у всех трех рассмотренных материалов превышает 3 эВ и что в случае укладывания этих трех материалов в виде сэндвича валентные зоны и зоны проводимости соседних слоев не соприкасаются. Отсюда следует, что общая зона проводимости получившегося сэндвича при отсутствии приложенного к внешним стенкам двойного барьера потенциала смещения будет пустой.

Предположим, что слева и справа от внешних слоев сэндвича к нему прилегают два плоских электрода, изготовленных из металла с уровнем электрохимического потенциала μ, расположенным ниже, чем уровень дна зоны проводимости у среднего слоя сэндвича ($E_{cbCF} = -4.45$ эВ). Таким металлом

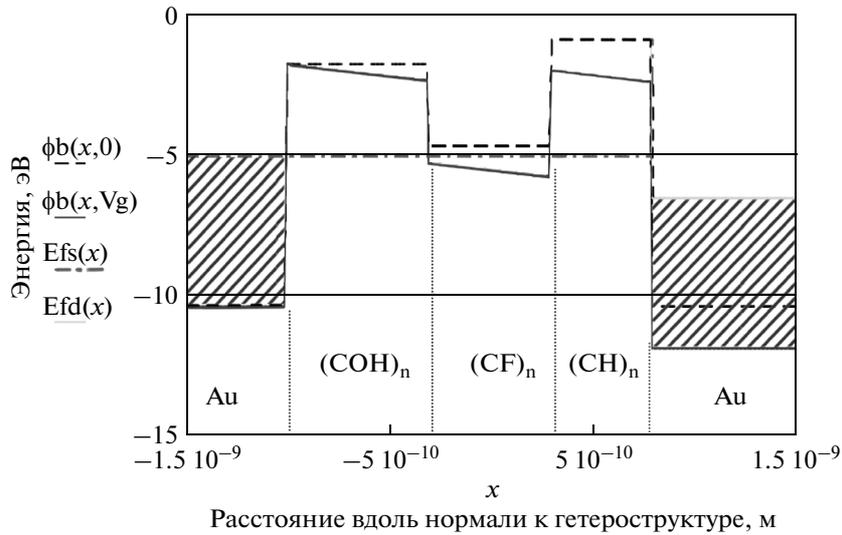

**Рис. 1.** Простейшая зонная диаграмма, описывающая двух барьерную туннельную структуру, построенную из трех слоев широкозонных 2D полупроводников — производных графена: пергидрокси графен $(COH)_n$, флюорографен $(CF)_n$ и графан $(CH)_n$, к которым по бокам прилегают две золотые пластинки с дном зоны проводимости на уровне $-10.5$ эВ и уровнем химического потенциала $-5.1$ эВ. Штриховая линия $\Phi_b(x, 0)$ — дно общей зоны проводимости в отсутствии потенциала смещения (дно квантовой ямы между барьерами выше уровня Ферми золота в левом электроде — штрих-пунктирная линия $Efs(x)$), сплошная тонкая ломанная линия $\Phi_b(x, V_g)$ — дно зоны проводимости при приложении потенциала смещения $V_g = 0.6$ В (дно квантовой ямы ниже уровня Ферми золота $Efs(x)$). Толстая сплошная линия $Efd(x)$ — уровень Ферми золота в правом электроде при подаче потенциала смещения. Потенциалы барьеров для простоты рассмотрения предполагаются трапецеидальными.

является, например, золото: $\mu_{Au} = -5.1$ эВ [13, 14]. Тогда будет выполнено условие $\mu_{Au} < E_{cbCF}$, т.е. туннельный ток через первый горбик двугорбого барьера в зону проводимости флюорографена, в отсутствии напряжения смещения на внешних металлических обкладках, отсутствует в силу закона сохранения энергии. Простейшая зонная диаграмма, описывающая двух барьерную структуру транзистора, соответствующую работе [9], но построенная из рассчитанных нами производных графена, изображена на рис. 1.

Заметим, что в классических кристаллических полупроводниках или в гетероструктурах на границах контактирующих слоев содержится большое количество поверхностных состояний (дефектов), которые смещают уровень Ферми внутри слоев, по сравнению с контактирующим с ними металлом. Мы в нашей модели имеем дело не с полированными кристаллами или выращенными методом MBE гетероструктурами, а с 2D полимерами (макромолекулами) — производными графена. Поэтому мы также, как и авторы, рассматривающие идеальные, имеющие гексагональную структуру, конструкции на основе пленок (макромолекул) графена и гексагонального двумерного BN [3–5], можем в своих построениях на уровне начальных моделей рассматривать наши пленки как три идеальные бездефектные макромолекулы: $(COH)_n$–$(CF)_n$–$(CH)_n$, И затем уже вносить диктуемые экспериментом уточнения, связанные с отличием реальных систем от модели.

При приложении к золотым электродам потенциала $V$, удовлетворяющего условию

$$0 < V < -2(E_{cbCF} - \mu_{Au})/e, \qquad (1)$$

где $e$ — заряд электрона, $V$ — потенциал смещения на внешних электродах, появится лишь малый туннельный ток через основание туннельного барьера из первого электрода на противоположный (затворный) электрод под дном зоны проводимости флюорографена. Однако, при приложении к внешним обкладкам разности потенциалов

$$V \geq -2(E_{cbCF} - \mu_{Au})/e, \qquad (2)$$

уровень Ферми левого электрода окажется выше, чем дно зоны проводимости слоя флюорографена и туннельный ток из первого золотого электрода будет проходить через тонкий первый туннельный барьер и попадать сначала в зону проводимости флюорографена $(CF)_n$, а затем уже проходить сквозь второй тонкий туннельный барьер. Заметим, что здесь, как и в работе [9], первый, по отношению к истоку, туннельный барьер (у нас это пергидрокси графен $(COH)_n$) имеет высоту немного меньшую, чем второй барьер (графан $(CH)_n$), за которым следует затворный электрод. Подобное соотношение высот барьеров, как и в работе [9], преследует своей целью увеличение полезного то-ка через первый барьер в квантовую яму и уменьшение паразитного тока через второй барьер на затворный электрод.

Для вычисления этих токов построим более точную модель двойного туннельного барьера. Для этого, согласно [15–17], необходимо учесть электрический потенциал сил "изображения" электрона на границе металл-диэлектрик. Потенциал сил "изображения", согласно [18] имеет вид $U(x) \approx -e^2/(4x 4\pi\varepsilon_0\varepsilon_r)$, где $x$ — расстояние от поверхности металла, $\varepsilon_0$ — электрическая постоянная вакуума, $\varepsilon_r$ — относительная диэлектрическая постоянная среды. Особенность этой функции при $x \to 0$, согласно [19], принято устранять, записывая потенциал сил "изображения" в виде $U(x) \approx -e^2/[(4x 4\pi\varepsilon_0\varepsilon_r) + e^2/E_b]$, где $E_b$ — уровень дна зоны проводимости металла.

Согласно нашим предыдущим квантово-химическим вычислениям системы энергетических уровней в так называемых "магических" золотых нано-кластерах $Au_{13}$, $Au_{55}$, $Au_{147}$ [20, 21], в них уже существует аналог зоны проводимости, т.е. система тесно расположенных частично занятых верхних уровней энергии, отделенных от ниже лежащих уровней запрещенной зоной, шириной ~50 эВ. Дно этой "зоны проводимости" в них располагается на уровнях энергии на 12.5 эВ, 11 эВ и 10.5 эВ ниже уровня вакуума, соответственно. Как видно из примера, эти значения, по мере увеличения размера золотого кластера, стремятся к асимптотическому значению ~10.5 эВ. По этой причине, в случае рассматриваемых нами золотых нано электродов, состоящих из $10^3$–$10^4$ атомов, в качестве уровня дна зоны проводимости возьмем значение $E_b = -10.5$ эВ.

На рис. 2 показано построение результирующего потенциала двойного туннельного барьера с помощью суммирования потенциала сил "изображения" и функции, описывающей форму дна зоны проводимости полупроводникового сэндвича, составленного из трех широко зонных полупроводников $(CH)_n - (CF)_n - (CH)_n$.

На рис. 2а жирной кривой показан ход дна этой общей зоны проводимости (распределение потенциальной энергии электрона), а также пунктирной кривой дно общей запрещенной зоны в случае объединения трех упомянутых материалов в трехслойный сэндвич $(CH)_n–(CF)_n–(CH)_n$ с учетом Ван дер Ваальсовского взаимодействия соприкасающихся слоев этого сэндвича. На рис. 2б показан потенциал сил "изображения" в промежутке между двумя плоскими золотыми электродами, разделенными промежутком, равным толщине сэндвича (1.8 нм) с учетом средней диэлектрической постоянной входящих в него слоев, которую мы приняли равной 2, (как у объемного тефлона $(CF)_n$ или у объемного полиэтилена $(CH)_т$) Как видно из рис. 2б, суммарный вклад в потенциальную энергию электрона "потенциала сил изображения" в центре промежутка между металлическими пластинами равен $\delta = -0.34$ эВ. На эту величину будет понижено дно квантовой ямы, которое образует дно зоны проводимости среднего слоя сэндвича — флюорографена $(CF)_n$. С учетом этой добавки формула (2) преобразуются к виду:

$$V \geq -2(E_{cbCF} + \delta - \mu_{Au})/e. \quad (2')$$

Подставив в эту формулу численные значения входящих величин, получим значение потенциала смещения, открывающего затвор:

$$V \geq 2 \times 0.31 = 0.62 \text{ В}. \quad (2'')$$

В результате суммирования графиков функций, изображенных на рис. 2а и 2б, получаем уточненную по сравнению с рис. 1 зонную диаграмму двойного туннельного барьера — рис. 2в.

## 2. ВЫВОД ФОРМУЛ ДЛЯ ПЛОТНОСТИ ТУННЕЛЬНЫХ ТОКОВ

Вычислим теперь токи, проходящие через первый и второй туннельные барьеры. В случае пространственной симметрии системы, дающей возможности разделить продольную (поперек барьера), и поперечную (параллельно барьеру) компоненты энергии электронов, плотность тока через тонкий диэлектрический туннельный барьер, приходящаяся на единицу площади дается формулой [18]:

$$J = \frac{2e}{h} \int_0^\infty dE[f_S(E) - f_D(E + eV)] \int \frac{d^2\vec{K}_\parallel}{(2\pi)^2} D(E, \vec{K}_\parallel), \quad (3)$$

где $e$ — заряд электрона, $h$ — постоянная Планка, $V$ — потенциал смещения, приложенного к туннельному барьеру, $f_S(E)$ и $f_D(E)$ — функции распределения Ферми–Дирака для истока и стока, соответственно. В нашем случае, согласно [18],

$$f_S(E) = f_D(E) = \{1 + \exp[(E - E_F)/k_B T]\}^{-1},$$

где $E_F = 5.5$ эВ — энергия Ферми золота [13,14] $k_B$ — постоянная Больцмана, $T$ — абсолютная температура $D(E, \vec{K}_\parallel)$ — прозрачность туннельного барьера. Интегрирование под знаком второго интеграла ведется по значениям волновым векторов электронов $\vec{K}_\parallel$, направленным параллельно к плоскости контакта.

В случае низких ($T = 300$ К) температур $k_B T = 0.026$ эВ $\ll |eV| = 0.6$ эВ, выражение (3), согласно [18], принимает следующий вид:

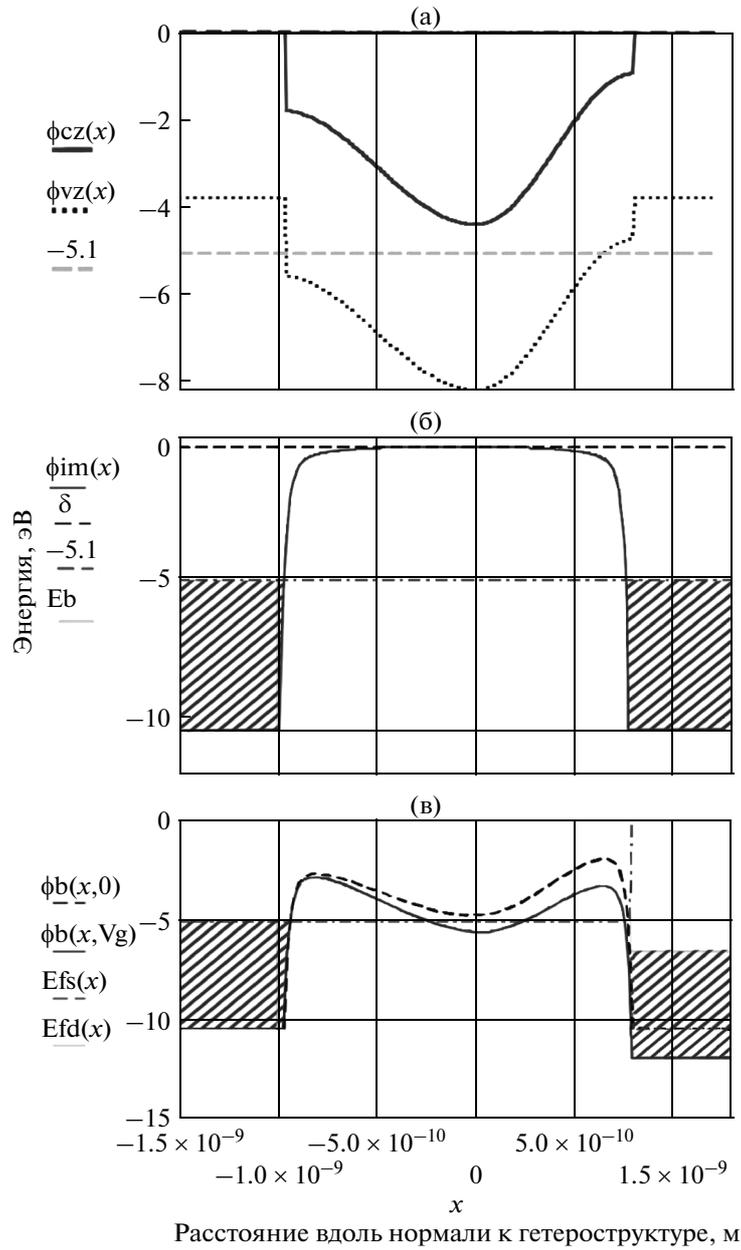

**Рис. 2.** Построение уточненной зонной диаграммы, описывающей двух барьерную туннельную структуру из трех слоев широкозонных 2D полупроводников – производных графена: $(COH)_n$, $(CF)_n$ и $(CH)_n$. Штриховая линия $\Phi_b(x, 0)$ – дно общей зоны проводимости в отсутствии потенциала смещения (дно квантовой ямы между барьерами выше уровня Ферми золота в левом электроде – штрих-пунктирная линия Efs($x$)), сплошная тонкая линия $\Phi_b(x, V_g)$ – дно зоны проводимости при приложении потенциала смещения $V_g = 0.6$ В (дно квантовой ямы ниже уровня Ферми золота Efs($x$)). Толстая сплошная линия Efd($x$) – уровень Ферми золота в правом электроде при подаче потенциала смещения. Потенциалы барьеров для простоты рассмотрения Потенциалы барьеров уже предполагаются деформированными Ван-дер Ваальсовским взаимодействием слоев в центре структуры и потенциалом сил "изображения" на внешних сторонах двойного барьера.

$$J = \frac{2e}{h} \int_{E_F+eV}^{E_F} dE \int \frac{d^2 K_{\|}}{(2\pi)^2} D\left(eV; E - \frac{\hbar^2 K_{\|}^2}{2m_e^*}\right), \quad (4)$$

где эффективная масса электрона в золоте равна $m_e^* = 1.1\, m_e$, согласно [14]. В случае малой по сравнению с энергией Ферми $E_F = 5.5$ эВ величиной энергии смещения $|eV|$ это выражение, согласно [18], можно преобразовать к виду:

$$J = \frac{2e^2}{h} \frac{m^* V}{\hbar^2} \int_0^{E_F} D(eV_g; E_F - E_{\|}) dE_{\|}. \quad (5)$$

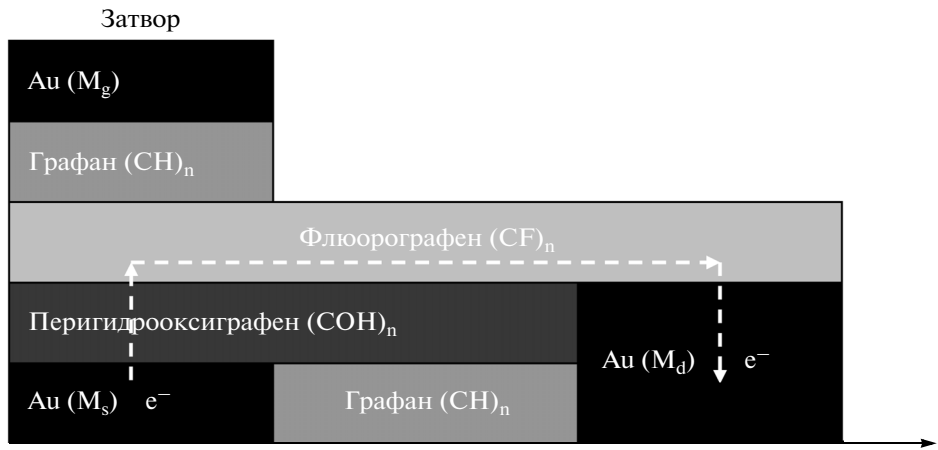

**Рис. 3.** Схематическое изображение двух барьерной транзисторной структуры в разрезе. Черным цветом закрашены металлические электроды (золото), светло серым — флюорографен, через который баллистически пролетают электроны, темно серым — пергидроокси графен, служащий первым барьером, средне серым — графан, служащий вторым барьером, а также уплотняющей подкладкой под слоем пергидро окси графена.

При прозрачности второго туннельного барьера много меньшей единицы, электроны будут отражаться от него назад, т.е. по направлению к первому барьеру. В свою очередь, коэффициент отражения первого туннельного барьера по отношению к этим электронам равен точно единице в силу принципа Паули. Т.о. при выполнении условия (2') электроны начнут накапливаться в зоне проводимости флюорографена, т. е. в квантовой яме и если их не отводить, то, как мы покажем в дальнейшем изложении, накопившийся заряд приведет к прекращению тока.

Рассмотрим способ отведения этого заряда. Предположим, что средний, состоящий из флюорографена, слой туннельной структуры $(COH)_n$–$(CF)_n$–$(CH)_n$ сделан шире основного сэндвича на полосу, с шириной, немного меньшей длины свободного пробега электронов в зоне проводимости флюорографена. Как известно [22], эта величина, так называемая баллистическая длина когерентности, равна $\xi_{Tb} \approx \hbar v_F / k_B T$, где $v_F$ — скорость электронов на уровне Ферми в 2DEG в флюорографене, равная $v_F = \sqrt{2E_{fCF}/m_e^*}$. Уровни Ферми в золотом истоковом электроде и в зоне проводимости флюорографена совпадают. Однако дно зоны проводимости флюорографена в открытом состоянии транзистора расположено на величину

$$\Delta = (E_{lumoCF} - \delta - \mu_{Au} + eV_g/2) \quad (6)$$

ниже уровня Ферми золота. Этой величине и будет равна энергия Ферми в квантовой яме $E_{fCF} = \Delta$. Значение продольной эффективной массы электрона $m_e^*$ в флюорографене в известных нам публикациях отсутствует. По этой причине можно либо, следуя работам [4, 5], считать ее подгоночным параметром, который будет точно определен в будущем эксперименте, либо положить ее равной эффективной массе электрона в графене $m_e^* = 0.19 m_e$ как в [4]. Подставляем эти значения энергии Ферми и эффективной массы в выше приведенные формулы. При потенциале смещения $V_g > 0.7$ В получим значение Фермиевской скорости $v_F > 2.1 \times 10^5$ м/с и значение баллистической длины когерентности в слое флюорографена $\xi_{T,b} > 35$ нм. На эту величину мы можем сделать лист флюорографена шире, чем остальные листы сэндвича. Предположим далее, что к внешней половине этой полосы прилегает золотой электрод шириной 10 нм. Пусть к этому электроду приложен потенциал $V_d = 1$. В относительно истока. С учетом этих предположений конфигурация нашего транзистора приобретет вид, изображенный на рис. 3.

Учитывая диэлектрические свойства флюорографена, можно предположить, что в участок этого широко зонного полупроводника, лежащий между основным сэндвичем, зажатым между двумя обкладками конденсатора, и стоком будет частично проникать электрическое поле, обусловленное существованием разных потенциалов на электродах истока и стока. Это поле будет дополнительно ускорять попавшие в зону проводимости флюорографена электроны, по направлению к стоку, еще больше увеличивая среднее значение скорости электронов и увеличивая тем самым баллистическую длину когерентности $\xi_{T,b}$.

Электроны, туннелировавшие через первый барьер, образованный пергидрокси графеном, в зону проводимости флюорографена, будут иметь два канала для дальнейшего распространения. Это а) канал баллистического прохождения вдоль тонкого слоя флюорографена на золотой стоковый электрод и б) канал, обусловленный туннельным переходом сквозь второй барьер, поперек слоя графана, на внешний (управляющий) золотой электрод, к которому приложено полное напряжение смещения (см. формулы (1, 2) Для оценки относительных и абсолютных величин токов, проходящих по этим двум каналам, введем сначала формулы для коэффициентов прозрачности соответствующих барьеров.

### 3. КОЭФФИЦИЕНТЫ ПРОЗРАЧНОСТИ КВАНТОВЫХ БАРЬЕРОВ

Для вычисления коэффициента прозрачности квантовых барьеров применим квазиклассическую формулу ВКБ [18]:

$$D_{i,k}(eV_g, E - E_\parallel) = \exp\left(-(2/\hbar)\left|\int_{x_i}^{x_k}\sqrt{2m_e^*(E - U(x, eV_g) - E_\parallel)}\,dx\right|\right). \quad (7)$$

Здесь $x_i$ и $x_k$ — классические точки поворота в пределах соответствующего туннельного барьера, в которых подкоренное выражение под символом интеграла меняет знак; $U(x, eV_g)$ — электростатическая потенциальная энергия внутри туннельного барьера, $V_g$ — управляющее напряжение на затворе, $m_e^*$ — поперечная по отношению к барьеру эффективная масса внутри барьера, которую, следуя работе [4], можно было бы положить подгоночным параметром, определяемым в эксперименте, но в силу малой толщины барьера мы пока принимаем ее равной массе электрона в вакууме, e — заряд электрона, $E$ — уровень продольной энергии, который электрон имеет перед данным барьером, $E_\parallel$ — компонента энергии электрона в направлении, параллельном барьеру [18].

Введем обозначения, учитывающие наличие двух последовательных барьеров: $D_{12}(eV_g; E - E_\parallel)$ для первого барьера, образованного слоем $(COH)_n$ толщины 0.7 нм и $D_{34}(eV_g; E - E_\parallel)$ для второго барьера, образованного слоем $(CH)_n$ толщины 0.5 нм, а также обозначение $D_{14}(eV_g; E - E_\parallel)$ для барьера суммарной толщины сэндвича $(COH)_n$–$(CF)_n$–$(CH)_n$, для случая туннелирования под дном зоны проводимости среднего слоя $(CF)_n$.

### 4. ВЫЧИСЛЕНИЕ ПЛОТНОСТИ ЭЛЕКТРОННОГО ТОКА, ПРОХОДЯЩЕГО ЧЕРЕЗ ДВУХ БАРЬЕРНУЮ СТРУКТУРУ

Сначала вычислим ток $i_{LR}$, проходящий сквозь барьер полной толщины 1.8 нм из левого золотого электрода в правый электрод. Этот ток определится выражением (5), где в случае открытого затвора, т.е. при $|eV_g| \geq 0.6$ эВ для величины коэффициента прозрачности барьера $D_{LR}$ имеет место соотношение

$$D_{LR}(eV_g; E - E_\parallel) = \\ = D_{1,2}(eV_g; E - E_\parallel) * D_{3,4}(eV_g; E - E_\parallel),$$

а в случае закрытого затвора, при

$$|eV_g| < 0.6 \text{ эВ}:$$

$$D_{LR}(eV_g; E - E_\perp) = D_{1,4}(eV_g; E - E_\parallel).$$

При этом мы считаем, что в случае открытого затвора, промежуток между двумя барьерами, доступный для классического движения, электроны проходят без изменения квадрата модуля амплитуды волновой функции [18].

Случай так называемого резонансного туннелирования, без какого либо затухания в двойном барьере, мы здесь не рассматриваем, поскольку барьеры имеют разную высоту и ширину.

Ток $i_{LM}$, проходящий сквозь один первый барьер в зону проводимости флюорографена при открытом затворе, т.е. при $V_g > 0.62$ В, будет вычисляться по формуле (5) при подстановке в нее прозрачности барьера в виде $D_{LM}(eV_g; E - E_\parallel) = D_{1,2}(eV_g; E - E_\parallel)$. А при закрытом затворе при $V_g < 0.62$ В ток полагается равным 0 в силу закона сохранения энергии, т.к. в этом случае уровень дна зоны проводимости флюорографена оказывается выше Уровня Ферми золота в истоке. В результате подстановки туннельных коэффициентов, вычисляемых по формуле (7) с потенциалом, изображенным на рис. 3в, в формулу для плотности тока (5) и умножения полученного результата на площадь туннельного контакта $S = 4.8 \times 10^2$ нм$^2$, получим графики туннельных токов, изображенные на рисунках рис. 4.

Из графиков видно, что при открытом затворе ($V_g > 0.62$ В) сквозной ток электронов сквозь три слоя 2D полупроводников (рис. 4(а)) будет ~ на 5 порядков меньше, чем ток из левого электрода сквозь один слой пергидроксиграфена в зону проводимости флюорографена (рис. 4(б)). Отсюда следует, что, если средний, флюорографеновый слой сделать немного шире, чем остальные слои сэндвича и подвести к нему специальный металлический стоковый электрод с постоянным

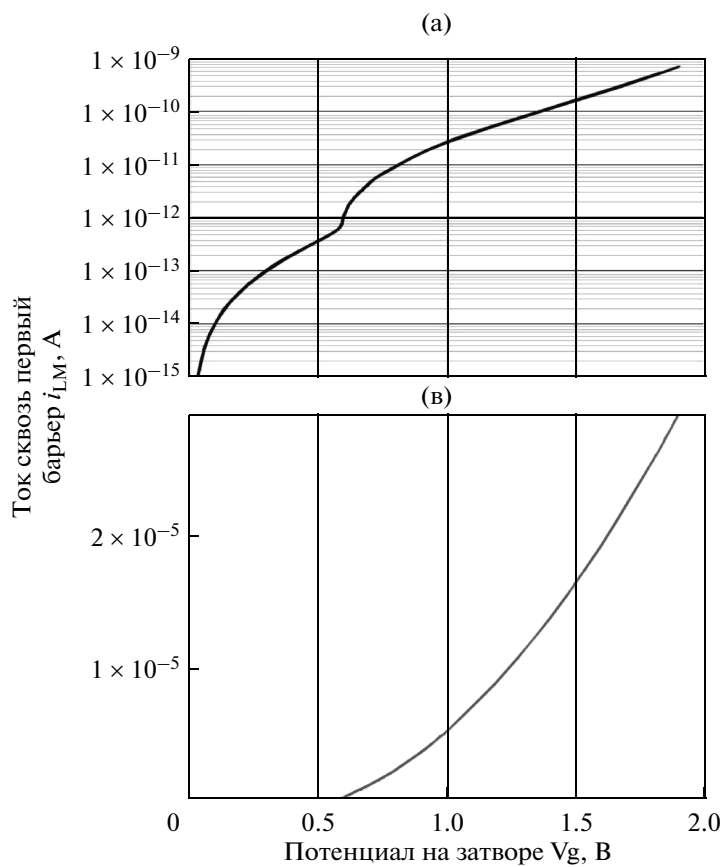

**Рис. 4.** ВАХ туннельных барьеров: *а* — ВАХ при туннелировании сквозь два барьера (в логарифмическом масштабе), *б* — ВАХ при туннелировании через первый барьер (в линейном масштабе). Точка излома на обоих графиках при $V_g = 0.6$ В соответствует открытию затвора, т.е. опусканию дна зоны проводимости флюорографена ниже уровня Ферми в первом электроде — истоке.

потенциалом, например $V_d = 1$ В, а ко второй обкладке конденсатора, являющейся затворным электродом получившейся системы, приложить потенциал открывания устройства 0.62 В (см. формулу (3)), то через систему пойдет ток, и полученное устройство будет работать как транзистор. Причем, ток, попадающий на стоковый электрод, примыкающий к среднему слою квантовой ямы $(CF)_n$, будет на несколько порядков больше сквозного тока.

## 5. МАКСИМАЛЬНЫЙ ТОК ВДОЛЬ КВАНТОВОЙ ЯМЫ

Электроны, прошедшие сквозь первый барьер попадают в квантовую яму, образованную двумя барьерами, т.е. в область, где существует двумерный электронный газ, так называемый 2DEG. Определим максимальную плотность электронного тока, который бы возникал в квантовой яме при контакте ее среднего слоя $(CF)_n$ с одной стороны непосредственно с массивным 3D проводником. Эта величина позволит нам дать оценку сверху для тока в квантовой яме в случае, если контакт с массивным 3D проводником осуществляется через туннельный барьер.

Для определения максимальной плотности электронного тока вдоль квантовой ямы, т.е. вдоль слоя флюорографена $(CF)_n$, необходимо применять баллистическую формулу [18], преобразованную к двумерному случаю. Формула (3) тогда примет вид:

$$J = \frac{2e}{h} \int dE [f_S(E) - f_D(E + eV)] \int \frac{d\vec{K}_\parallel}{(2\pi)} D(E, \vec{K}_\parallel), \qquad (8)$$

где $E_\parallel = \dfrac{\hbar^2 k_\parallel^2}{2m_e^*}$. При баллистическом распространении электронов вдоль квантовой ямы туннельный коэффициент прозрачности $D(eV; E - E_\parallel) \cong 1$, предполагая по прежнему $k_B T \ll eV$, формулу (10) можно преобразовать к виду

$$J = \frac{2e\sqrt{2m_e^*}}{h^2} \int\limits_{E_F - eV}^{E_F} dE_S \int\limits_0^{E_s} \frac{dE_\parallel}{\sqrt{E_\parallel}}. \tag{9}$$

В этом случае интегралы берутся в конечном виде и мы получаем выражение для плотности тока в одномерном сечении двумерной квантовой ямы в виде:

$$J = \frac{2e\sqrt{2m_e^*}}{h^2}\left(\frac{2}{3}\right)\left[E_{FS}^{3/2} - (E_{FS} + eV_D)^{3/2}\right], \tag{10}$$

Где $V_d$ — потенциал смещения на стоковом металлическом электроде, примыкающем к листу флюорографена, $m_e^* = 0.19 m_e$ — значение продольной эффективной массы электрона в слое флюорографена, которое мы приняли согласно [4]. Для получения искомого максимально возможного тока в нашей двумерной квантовой яме, это выражение необходимо умножить на длину ее одномерного сечения, т.е. на длину стороны квадрата нашего сэндвича $a = 2.2 \times 10^{-8}$ м.

$$I_{\max} = \frac{2e\sqrt{2m_e^*}}{h^2}\left(\frac{2}{3}\right)\left[E_{FS}^{3/2} - (E_{FS} + eV_D)^{3/2}\right]a. \tag{11}$$

При подстановке значений констант и характерных для рассматриваемой системы величин: $E_{FS} = E_{fAu} = 5.5$ эВ, $V_D = 1$ В в формулу (13), получим $I_{\max 1} = 2.5 \times 10^{-3}$ А.

Для сравнения проведем оценку максимального тока вдоль слоя (полосы) флюорографена, шириной 22 нм, используя квазиклассическую формулу Шарвина [23, 24] для баллистической проводимости вдоль флюорографенового листка шириной $a = 22$ нм и толщиной $b = 0.6$ нм. Эта проводимость, согласно Шарвину [23, 24], равна $G_{Sh} = \dfrac{2e^2}{h} N$, где $N$ — число каналов, доступных для баллистического прохождения электронов: $N = (ab)/\lambda_{FCF}^2 \approx 49$ и $\lambda_{FCF} = h/\sqrt{2m_e^* E_{fCF}}$ — Фермиевская длина волны электрона в зоне проводимости флюорографена, а величина $E_{fCF} = \Delta$ определяется по формуле (6). Отсюда $I_{\max 2} = G_{Sh} V_d = \dfrac{2e^2}{h} N V_d \approx 3.8 \times 10^{-3} A$. Таким образом, мы разными методами получили для максимального тока оценки, отличающиеся ~ в 1.5 раза. Отсюда следует, что полученная по баллистической формуле с учетом эффекта ослабления тока за счет прохождения электронов сквозь первый туннельный барьер величина тока является оценкой снизу, и не будет уменьшаться за счет протекания электронов по двумерному каналу, т.е. этот ток может быть значительно (примерно на порядок) увеличен, если увеличить прозрачность первого туннельного барьера, за счет уменьшения его толщины, например, применив вместо пергидро окси графена с толщиной 0.7 нм гексагональный нитрид бора $(BN)_n$ с толщиной порядка 0.4 нм. В условиях нашей задачи, в соответствии с вычислениями по формуле (5), получаем для максимальной плотности тока, отнесенной к длине затвора значения, приведенные на графике рис. 5. Эти значения примерно соответствуют данным, приведенным в работах [1, 2, 4, 5] для полевых туннельных транзисторов на основе кремния и графена.

Поскольку, как мы установили выше, ток в нашей конструкции транзистора не будет уменьшаться в баллистическом канале стока по сравнению с тем значением, которое определяется прозрачностью первого туннельного барьера, в такой конструкции будет более четко проявляться насыщение тока открытого состояния транзистора. Это обстоятельство иллюстрируется на графиках, приведенных на рис. 6.

Из этих графиков видно, что на начальном участке ВАХ линейны, что соответствует выполнению закона Ома при баллистическом прохождении электронами флюорографенового листка вплоть до выхода графиков на значения плотно-

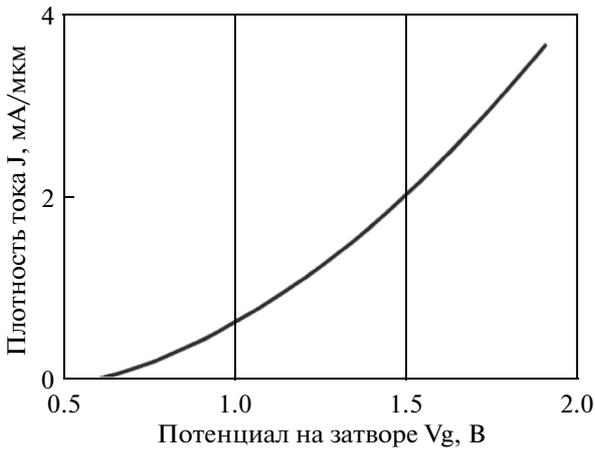

**Рис. 5.** Графики максимальной плотности тока от потенциала на затворе $V_g$.

сти тока, соответствующие насыщению, зависящие от разного пропускания первого туннельного барьера при разном напряжении $V_g$, заданном на затворе.

## 6. ЭЛЕКТРИЧЕСКАЯ КВАНТОВАЯ ЕМКОСТЬ И ОЦЕНКА СВЕРХУ БЫСТРОДЕЙСТВИЯ НАНОТРАНЗИСТОРА

Быстродействие в нашей модели транзистора на квантовой яме оценим, следуя подходу [9], при котором в рассмотрение была введена квантовая емкость. Согласно [9], именно большая величина квантовой емкости двумерного электронного газа (2DEG) квантовой ямы, по сравнению с величиной обычной, так называемой "геометрической", емкости обычных металлических проводников является основным фактором, определяющим быстродействие в рассматриваемой модели транзистора. На рис. 7 изображена, заимствованная нами из [9], схема соединения обычных и квантовых емкостей, эквивалентная нашей модели нанотранзистора.

Здесь $C_1$ и $C_2$ обозначают обычные "геометрические" емкости, которые соответствуют плоским конденсаторам, одной из обкладок которых является обычный металлический электрод истока или затвора, соответственно, а второй обкладкой является "2DEG" квантовой ямы. Емкости $C_1$ и $C_2$ вычисляются по обычной формуле: $C_1 = C_2 = \varepsilon_0 \varepsilon S/d$, где $\varepsilon_0$, — абсолютная и $\varepsilon r = 2$ относительная диэлектрическая постоянная, соответственно, $S = a^2 = 4.84 \times 10^{-16}$ м$^2$ — одинаковая площадь пластин электродов истока и затвора, $d \approx 0.9 \times 10^{-9}$ м — расстояния между металлической пластиной и средней плоскостью квантовой ямы, соответствующие внешним металлическим электродам (истоку и затвору). Величина $C_q$ квантовой емкости, находящейся между туннельными барьерами, окружающими квантовую яму, согласно [9], вычисляется по формуле $C_q = e^2 D_q S$, где $D_q = g_v m_e^*/(\pi \hbar^2)$ — плотность электронных состояний в 2DEG — системе, $S$ — ее площадь, $e$ — заряд электрона, $\hbar$ — постоянная Планка, деленная на $2\pi$. Значения $g_v$ — так называемого долинного фактора вырождения и $m_e^*$ — эффективной массы электрона в квантовой яме берем, за отсутствием в литературе данных для (CF)$_n$, равными $g_v = 4$ и $m_e^* = 0.19 m_e$, как у гра-

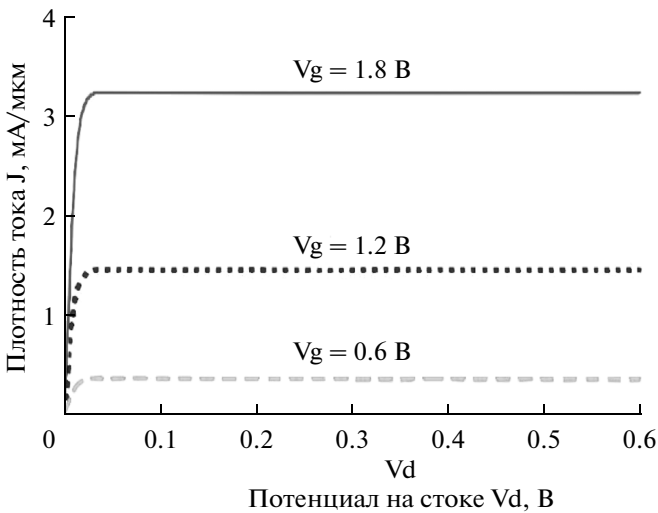

**Рис. 6.** Графики ВАХ, иллюстрирующие насыщение плотности тока открытого состояния при разных потенциалах на затворе.

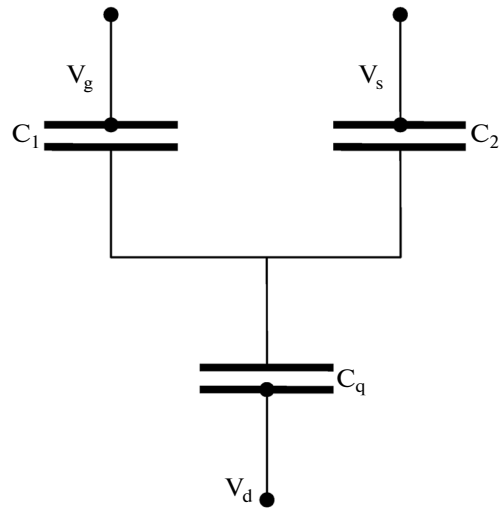

**Рис. 7.** Схема соединения конденсаторов эквивалентных нашей модели транзистора.

фена, согласно [4, 5]. При вычислении величин емкостей принимаем величину относительной диэлектрической постоянной ε$r$ = 2, как у объемного тефлона. В результате вычислений получаем значения: $C_q = 2.5 \times 10^{-16}$ Ф, $C_1 = C_2 \approx 10^{-17}$ Ф. Т.о. величина квантовой емкости квантовой ямы ~ в 25 раз больше, чем "геометрическая" емкость каждого из электродов. Полная емкость изображенного на рис. 7 составного конденсатора, эквивалентного нашей системе (нанотранзистору), вычисляется, как $C_\Sigma = 1/[1/C_q + 1/(C_1 + C_2)]$. Она равна $C_\Sigma \approx 1.8 \times 10^{-17}$ Ф. Постоянная времени разрядки этого эквивалентного конденсатора τ вычисляется как $\tau = R_q C_\Sigma$, где $R_q = (G_{Sh})^{-1}$ и равна τ = $= 4.7 \times 10^{-15}$ с. Таким образом электрическая емкость квантовой ямы в нашей модели транзистора не ограничивает его рабочей частоты, вплоть до $10^{14}$ Гц. Представляет интерес вычислить максимальный электрический заряд, способный разместиться внутри квантовой ямы. Этот заряд равен $\Delta Q_{qw} = eD_q SE_{fCF} \approx -3.6 \times 10^{-16}$ Кл и соответствующий этому заряду максимальный ток $I_{max3} =$ $= \Delta Q_{qw}/\tau = 5.4 \times 10^{-3} A$, а также соответствующее этому заряду изменение потенциала квантовой ямы

$$\Delta \Phi_{qw} = \Delta Q_{qw}/C_q = \frac{eD_q SE_{fCF}}{e^2 D_q S} = E_{fCF}/e,$$

которое, как и следовало ожидать, по закону сохранения энергии стремится компенсировать действие внешнего потенциала, открывающего затвор транзистора, см. формулу (6). Т.о. как мы указывали в предыдущем параграфе, в отсутствии стока электронов из квантовой ямы произойдет полное запирание транзистора. Время, необходимое для накопление такого заряда в отсутствии стока равно $\Delta Q_{qw}/I_{LM} \approx 3.6 \times 10^{-16}/2 \times 10^{-5} = 1.8 \times 10^{-11}$ с, что ~ в 20 больше обратной частоты транзистора терагерцового диапазона. Поэтому заряд, накапливающийся в течение рабочего периода $10^{-12}$ с, даже в отсутствии стока, не будет существенно ограничивать ток такого транзистора.

## ЗАКЛЮЧЕНИЕ

• Проведены квантово-химические расчеты зонной структуры трех широкозонных 2D полупроводников-полимеров: флюорографена $(CF)_n$, графана $(CH)_n$ и пергидрокси графена $(COH)_n$. Получены значения уровней энергии: дна зоны проводимости $E_{cb}$: $E_{cbCOH} = -1.8$ эВ, $E_{cbCF} = -4.45$ эВ, $E_{cbCH} = -0.9$ эВ и потолка валентной зоны $E_{vb}$: $E_{vbCOH} = -4.985$ эВ, $E_{vbCF} = -7.728$ эВ, $E_{vbCF} = -4.762$ эВ. Сэндвич, составленный из этих трех слоев – Ван дер Ваальсовская гетероструктура:

$(COH)_n–(CF)_n–(CH)_n$ будет представлять собой двух барьерную структуру с пустой общей зоной проводимости. Этот сэндвич, помещенный между двумя слоями металла (золота), со значением уровня химического потенциала μ более низким, чем наиболее низко расположенное дно зоны проводимости (у среднего слоя сэндвича $(CF)_n$), т.е. $\mu_{Au} = -5.1$ эВ $< E_{cbCF} = -4.45$ эВ при приложении к внешним слоям металла разности потенциалов $\Delta V_g \geq 2(\mu_{Au} - E_{cbCF}) = 0.62$ В будет заполняться электронами и превратится в квантовую яму.

• На основе этих квантово-химических расчетов предложена и далее рассчитана модель нанотранзистора, с полевым затвором на туннельном переходе из внешнего металлического электрода в квантовую яму, образованную трехслойной Ван дер Ваальсовской гетероструктурой, с быстродействием до $10^{12}$ Гц.

• "Затворная" часть нанотранзистора (см. рис. 3) представляет собой квадратный сэндвич со стороной $a \sim 20$ нм из 5 слоев различных материалов, каждый слой толщиной в диапазоне ~0.5–2 нм, и общей толщиной ~5 нм, а именно: $M_S–(COH)_n–(CF)_{nD}–(CH)_n–M_G$. Эти слои включают в себя 2 металлических электрода: 1) истоковый электрод $M_S$ снизу сэндвича, 2) $M_G$ – электрод, управляющий затвором, сверху сэндвича. Средний, образующий дно квантовой ямы слой сэндвича $(CF)_{nD}$, ~ на 35 нм шире остальных и является "стоковым". К внешней половине этой полосы $(CF)_{nD}$, шириной ~10 нм примыкает третий (стоковый) металлический электрод $(M)_D$, находящийся под постоянным потенциалом и не входящий в общий "затворный" сэндвич.

• Показано, что порог открытия затвора составляет $V_g \sim 0.62$ В по отношению к истоку и диапазон изменения полного тока составит $0–2 \times 10^{-5}$ А при изменении потенциала на затворе в интервале (0.62–1.0) В.

• Ток закрытого состояния при $V_g < 0.62$ В равен 0. Таким образом отношение тока открытого состояния к току закрытого равно ∞.

• Ток открытого состояния быстро (при потенциале на стоке порядка $V_D = 0.05$ В) достигает насыщения и выходит на постоянное плато (см. рис. 6).

• Паразитный ток утечки через затворный электрод на 5 порядков меньше рабочего тока с истока на стоковый электрод.

• Согласно А. Гейму [6], создание гетероструктур, подобных предложенным в данной модели нанотрантранзистора, за счет настилания слоев широко зонных 2D полупроводников при переходе на литографию с разрешением лучше 22 нм будет технологичнее, чем у напыляемых МОП нанотранзисторов.

• Следует отметить, что обсуждаемый в статье новый материал, пергидрокси графен, еще не синтезирован и взят нами в качестве модели широко зонного 2D полупроводника, имеющего необходимые характеристики, а именно, уровень дна зоны проводимости значительно выше, чем у флюорографена и, одновременно, немного ниже, чем у графана. Это позволяет провести необходимые расчеты и смоделировать транзистор, в котором паразитный сквозной ток значительно (на 5 порядков) меньше рабочего тока. При использовании графана для обоих внешних слоев рассмотренного сэндвича эта разница составила бы всего 3 порядка. Пергидрокси графен мог бы быть синтезирован, например, при контакте графена с водяным паром, облучаемым ультрафиолетовым излучением.

• Предварительные расчеты показывают, что в качестве альтернативного материала, вместо пергидрокси графена, так же мог бы использоваться двумерный гексагональный нитрид бора BN. При этом рабочий ток должен возрасти примерно на порядок. Однако, точные расчеты требуют больших затрат машинного времени и станут предметом рассмотрения в нашей следующей работе.